\documentclass[twocolumn,showpacs,preprintnumbers,amsmath,amssymb,prl]{revtex4}
\usepackage{graphicx}
\usepackage{subfigure}
\usepackage{dcolumn}
\usepackage{bm}
\usepackage{epstopdf}
\begin{document}

\title{Evidence of mixed phases and percolation at the metal-insulator transition in two dimensions.}
\author{Shiqi Li}
\affiliation{Department of Physics, City College of New York, 160 Convent Ave., New York, New York 10031, USA\\
CUNY Graduate Center, 365 Fifth Avenue, New York, New York 10016,
USA}
\author{Qing Zhang}
\affiliation{Department of Physics, City College of New York, 160 Convent Ave., New York, New York 10031, USA\\
CUNY Graduate Center, 365 Fifth Avenue, New York, New York 10016,
USA}
\author{Pouyan Ghaemi}
\affiliation{Department of Physics, City College of New York, 160 Convent Ave., New York, New York 10031, USA\\
CUNY Graduate Center, 365 Fifth Avenue, New York, New York 10016,
USA}
\author{M. P. Sarachik}
\affiliation{Department of Physics, City College of New York, 160 Convent Ave., New York, New York 10031, USA\\
CUNY Graduate Center, 365 Fifth Avenue, New York, New York 10016,
USA}

\date{\today}

\begin{abstract}
The in-plane magnetoconductance of the strongly interacting
two-dimensional electron system in a silicon MOSFET
(metal-oxide-semiconductor-field-effect transistor) exhibits an
unmistakeable kink at a well-defined electron density, $n_k$. The
kink at $n_k$ is near, but not at the critical density $n_c$
determined from resistivity measurements, and the density at which
$n_k$ occurs varies with temperature. These features are
inconsistent with expectations for a quantum phase transition. We
suggest instead that this is a percolation transition and present a
detailed model based on the formation of a mixed insulating and
metallic phase within which a metal-insulator transition takes place
by percolation.
\end{abstract}

\maketitle

\section{Introduction}

Based on the famous 1979 paper by Abrahams {\it et al.}
\cite{Abrahams79}, as well as several carefully executed experiments
on different materials, it was assumed for many years that a
metal-insulator transition cannot occur and no metallic phase can
exist in a two-dimensional electron/hole system. It was therefore a
surprise when experimental studies in the 1990's appeared to show
that such a transition does take place in low disorder, dilute
two-dimensional (2D) electron systems when strong electron
interactions rather than the kinetic energy are dominant and
determine the behavior of the system \cite{Krav94,Krav95,Krav96}. In
addition to the dramatic change in resistivity that signals the
onset of a conducting phase, unusually interesting behavior was
found for the magnetoresistance both above and below the critical
electron density $n_c$. On both sides of the transition, the
resistivity rises sharply with increasing in-plane magnetic field up
to a field $B_{sat}$, above which it becomes essentially constant.
Shubnikov-de Haas measurements
\cite{okamoto99,vitkalov2000,vitkalov2001a} have demonstrated that
$B_{sat}$ signals the onset of full polarization of the electron
spins.

A great deal of discussion ensued concerning these experimental
observations: is this a true quantum phase metal-insulator
transition, which many believed cannot occur in two dimensions, or
can be explained within any one of a number of relatively benign
scenarios. (For reviews, see Refs. \cite{review1,review2,review3}.)

While abrupt changes and divergences have been reported at $n_c$ for a variety of physical properties \cite{review1,review2,review3,Mokashi2012}, the benign behavior of the magnetoresistance at the critical electron density has been an enigma since its discovery \cite{simonian97,pudalov97}. While the resistivity displays a sharp change as one crosses the transition at the critical density, the magnetoresistance appears to vary smoothly without exhibiting any change that would signal the onset of a new phase \cite{mertes99}.

In this paper we report the results of detailed measurements of the
magnetoresistance of a two-dimensional low-disorder, dilute,
strongly interacting system of electrons in a silicon
metal-oxide-semiconductor-field-effect transistor (MOSFET) over a
broad range of electron densities. Despite the striking similarity
of the magnetoconductance in the metallic and insulating phases and
the gradual evolution of the behavior found to date as the
transition is crossed, we report that the in-plane field $B_{sat}$
required to fully polarize the electron spins exhibits an abrupt
kink as a function of electron density that signals the occurrence
of a transition at a well-defined density $n_k$ that varies with
temperature and is near, but not at, the density $n_c$ for the
metal-insulator transition determined from resistivity measurements.
These features are inconsistent with the behavior expected for a
quantum critical transition. We propose instead that the kink
signals the occurrence of a transition by percolation within a mixed
metallic-insulating phase.

\section{Experimental procedure}

\begin{figure}[h]
\centering
\hspace{-.85in}
\includegraphics[width=0.6\textwidth]{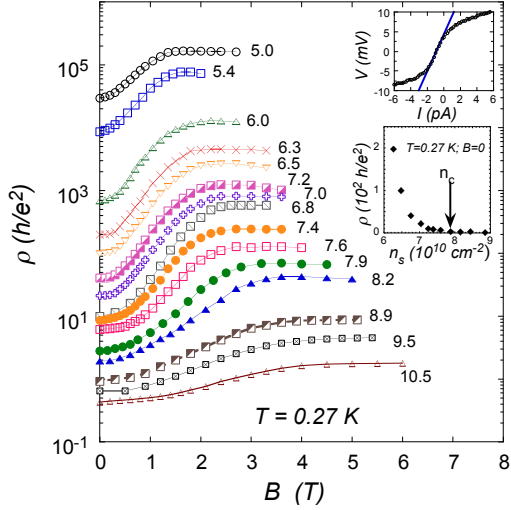}
\caption{\label{fig:Fig1} The resistivity as a function of parallel
magnetic field at $T = 0.27$ K plotted on a semilogarithmic scale
for different electron densities (in units of $10^{10}$ cm$^{-2}$),
as labeled. The small decrease of the resistance in some of the
curves for magnetic fields above saturation is due to minor
misalignment of the field away from the in-plane direction. Upper
Inset: The resistivity deduced from the slope of the linear portion
of the corresponding $I$-$V$ characteristic for each density and
temperature; here $T = 0.27$ K, $B=0$ T; $n_s \approx 4.7 \times
10^{10}$ cm$^{-2}$. Lower inset: A plot of $\rho$ as a function of
electron density $n_s$ illustrates the abrupt change in resistivity
at $n_c$.}
\end{figure}

Measurements were performed down to $0.27$ K in an Oxford Heliox He-3 refrigerator on
the same type of high-mobility silicon MOSFET samples as those used
in previous studies \cite{Mokashi2012, Shiqi2017}. Here we report data taken for a sample with  critical density $n_c \approx 7.74 \times 10^{10}$
 cm$^{-2}$ in the absence of magnetic field and $\approx 9.0 \times 10^{10}$
 cm$^{-2}$ in a field sufficient to fully polarize the electron spins \cite{Shiqi2017}. Contact resistance was
minimized by using a split-gate geometry which permits high electron
density to be maintained near the contacts independently of the
value of the electron density in the main part of the sample. This
is a particularly important feature that enables reliable
measurements in the dilute 2D electron system in the deeply
insulating state where the resistivity reaches very high values. By
contrast with the lock-in techniques that were sufficient for our
earlier measurements of higher density, metallic-like samples that
have relatively low resistivities \cite{Vitkalov2001b}, the
measurements here were taken by a protocol similar to that described
in Ref.~\cite{Shiqi2017}, where a Keithley Source Measure Unit SMU
236 was used to apply a small dc current (as low as a few pA)
through the sample and the voltage was measured by a standard
four-probe method. For each density and temperature, the resistivity
$\rho$ was deduced from the slope of the linear portion of the
corresponding $I$-$V$ characteristic, as shown in the upper inset to
Fig.~\ref{fig:Fig1}.

\section{Experimental results}

\begin{figure}[h]
\centering
\vspace{-.2in}
\includegraphics[width=0.65\textwidth]{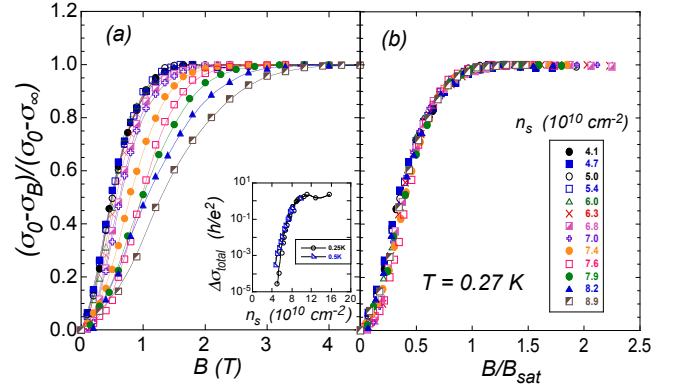}
\vspace{-1in} \caption{\label{fig:Fig2} (a) Normalized conductivity
as defined by Eq. (\ref{eqn:theratio}) plotted as a function of
in-plane magnetic field $B$;  $T=0.27$ K. The inset shows $\Delta
\sigma_ {total} = [\sigma(B=0) - \sigma(B\rightarrow\infty)]$ versus
electron density $n_s$. (b) Normalized conductivity as a function of
$B/B_{sat}$, where the fitting parameter $B_{sat}$ was chosen to
yield a collapse of the normalized conductivity onto a single
curve.}
\end{figure}

For various different electron densities, Fig.~\ref{fig:Fig1} shows
the resistivity at $T=0.27$ K as a function of magnetic field
applied parallel to the plane of the sample. In agreement with data
shown in earlier reports \cite{mertes99}, the in-plane
magnetoresistance rises dramatically with increasing magnetic field
and reaches a plateau above a density-dependent field $B_{sat}$. The
behavior of the magnetoresistance is qualitatively the same in the
insulating phase as it is in the conducting phase and appears to
evolve continuously and smoothly, with no indication that a
transition has been crossed. As shown in the lower inset of
Fig.~\ref{fig:Fig1}, this is in clear contrast with the dramatic
change found for the zero-field resistance as the electron density
is reduced below $n_c$, a change that becomes sharper and more
pronounced as the temperature is reduced into the milli kelvin
range.

\begin{figure}[h]
\centering
\vspace{-.6in}
\hspace{-.4in}
\includegraphics[width=0.7\textwidth]{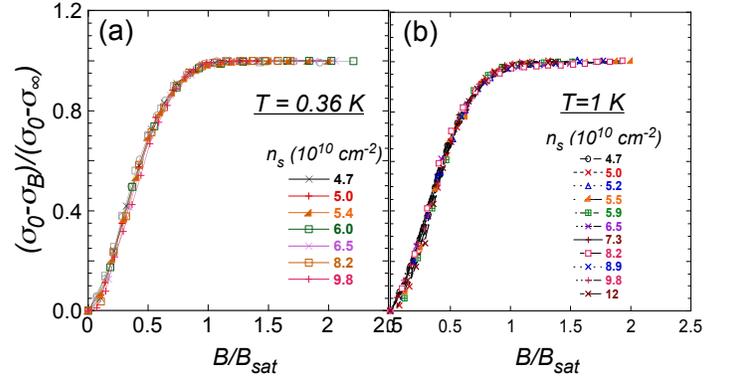}
\vspace{-1.4in} \caption{\label{fig:Fig3} To illustrate the collapse
obtained at each (constant) temperature, the normalized conductivity
is shown as a function of $B/B_{sat}$ at two temperatures above
base: (a) $0.36$ K; (b) $1.0$ K.}
\end{figure}

\begin{figure}[h]
\centering
\vspace{-.1in}
\hspace{-.4in}
\includegraphics[width=0.55\textwidth]{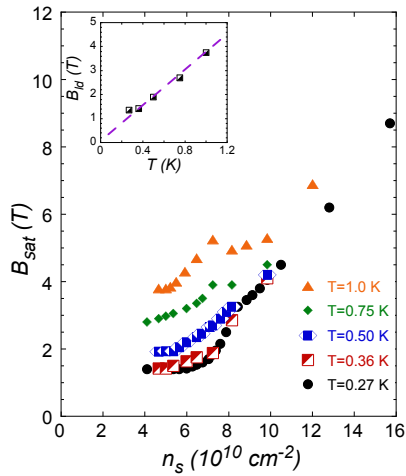}
\vspace{-.3in}
\caption{\label{fig:Fig4} $B_{sat}$ as a function of electron density for different temperatures, as labeled. The inset shows the terminal, low-density value of $B_{sat}$ as a function of temperature.}
\end{figure}

\begin{figure}[h]
\centering
\vspace{-.1in}
\hspace{-.4in}
\includegraphics[width=0.55\textwidth]{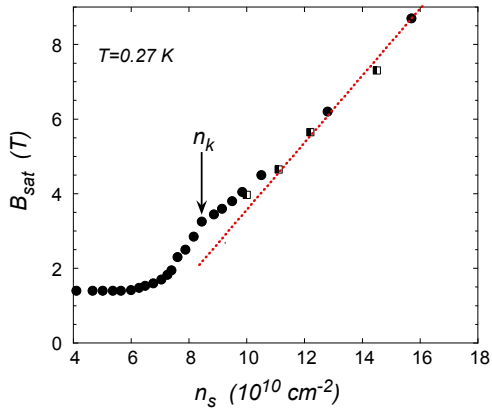}
\vspace{-.3in}
\caption{\label{fig:Fig5} $B_{sat}$ as a function of $n_s$ at base temperature. The closed circles are for the sample for which most of the data were obtained, while the half-open squares were obtained for a similar sample with a different value of $n_c$ (see the Appendix). To aid the discussion in the text, the dotted line is an approximate straight line fit to the high density values, extended to lower densities.}
\end{figure}

Following the procedure used in a previous study \cite{Vitkalov2001b}, we determine the normalized
magnetoconductivity:
\begin{equation} \label{eqn:theratio}
\sigma_{\textrm{norm}} \equiv \frac{\sigma(B=0) -
\sigma(B)}{\sigma{(B=0)} -
\sigma{(B\rightarrow\infty)}}=  \frac { \Delta \sigma} {\Delta \sigma_{total}}~.
\end{equation}
 Note that the normalized magnetoconductivity
 is simply the field dependent contribution to the
conductivity,
$\Delta \sigma = [\sigma(B=0) - \sigma(B)]$, normalized by its full value, $\Delta \sigma_ {total} = [\sigma(B=0)
- \sigma(B\rightarrow\infty)]$.

Figure~\ref{fig:Fig2} (a) shows the normalized conductivity at  $T=0.27$ K as a function of in-plane magnetic field for different electron densities. The inset is a plot of $\Delta \sigma_ {total} = [\sigma(B=0)
- \sigma(B\rightarrow\infty)]$ as a function of electron density. Notwithstanding the apparent continuity of the behavior of the magnetoconductance across the transition, the inset shows that there is a sharp change in the magnetoconductivity in the vicinity of a critical density $n_c$.

As shown in Fig.~\ref{fig:Fig2}(b), a density-dependent parameter
can be chosen that provides a collapse onto a single curve for all
the data obtained at base temperature; note that the scale of
$B_{sat}$ was chosen such that the onset of saturation occurs at
$x=1=B/B_{sat}$. Data were also taken at $0.36$, $0.50$, $0.75$ and
$1.0$ K. Figure~\ref{fig:Fig3} illustrates that the $B_{sat}$
obtained at $0.36$ and $1.0$ K are also self-similar and collapse
onto a single curve (data for $0.5$ and $0.75$ K not shown).

Figure~\ref{fig:Fig4} shows a plot of $B_{sat}$ for a broad range of electron density at five different low temperatures. A clear kink occurs at a density (that varies with temperature), below which $B_{sat}$ decreases with decreasing electron density and assumes a constant terminal value, $B_{ld}$. The inset shows a plot of the constant terminal value of $B_{ld}$ as a function of temperature. To aid the discussion below, $B_{sat}$ at the base temperature ($T=0.27$ K) is shown separately in Fig.~\ref{fig:Fig5}.

\begin{figure}[h]
\centering
\hspace{-0.4in}
\includegraphics[width=0.55\textwidth]{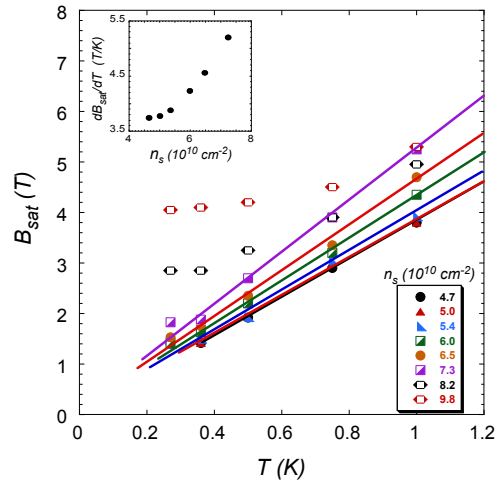}
\caption{\label{fig:Fig6} $B_{sat}$ as a
function of temperature $T$. For the six lower densities, which are insulating, the lines are linear fits to the data not including the data at the lowest temperature $T=0.27$ K, which lie above the curve. The upper two curves are on the metallic side of the transition. The inset shows the slope $A$ of the $B_{sat}$ vs. $T$ curves in the insulating phase as a function of electron density.
}
\end{figure}

Figure~\ref{fig:Fig6} shows $B_{sat}$ as a function of the
temperature $T$ for various different electron densities. For each
of the six densities on the insulating side of the transition,
$B_{sat}$ is consistent with the linear fits shown Thus, $B_{sat} =
AT$. Note that the lowest temperature points all deviate upward from
the straight lines. Although we ascertained that the resistance of
the sample exhibited exponential (variable-range-hopping) behavior
down to base temperature, it is possible that the electron system
was not  thermally well connected to the bath at our lowest
temperatures. More densely spaced data taken at lower temperatures
are needed to determine whether $B_{sat}$ tends to zero in the limit
of zero temperature. By contrast, $B_{sat}$ extrapolates to a finite
value at $T=0$ for the two highest (metallic) densities, signaling
the entry into a different phase. For the insulating phase, the
inset shows that the slope $A$ of the straight line fits as a
function of electron density decreases as the density is reduced and
levels off to a constant value.

\section{Discussion}

In this paper we report measurements of the resistivity as a
function of in-plane magnetic field in the strongly interacting
two-dimensional electron system in a silicon MOSFET over a broad
range of electron densities spanning the insulating and metallic
phases at several temperatures below $1$ K. Plotted as a function of
in-plane magnetic field, $B$,  the normalized conductivity defined
by Eq. (\ref{eqn:theratio}) then yields a set of self-similar curves
that can be collapsed onto a single curve for all measured electron
densities and temperatures using a parameter $B_{sat}$, where
$B_{sat}$ is the in-plane magnetic field required to fully polarize
the electron spins in the system. $B_{sat}$ is found to exhibit
surprising, complex and interesting behavior as shown in
Figs.~\ref{fig:Fig4} and \ref{fig:Fig5}. We call attention to the
following notable features:

(A) While $B_{sat}$ follows the straight line behavior expected for a Fermi liquid at high electron densities (see the dotted red line in Fig.~\ref{fig:Fig5}), there is a clear upward deviation from the straight line as the electron density approaches a kink.

(B) Shown by the arrow in Fig.~\ref{fig:Fig5}, there is an
unmistakable kink at a well-defined density $n_k$ followed by an
abrupt decrease of $B_{sat}$ as the electron density is further
reduced. The kink at $n_k$ signals the occurrence of a transition at
a density that coincides with neither the zero-field critical
density $n_c$ nor its value in a magnetic field.  It occurs at
finite temperature and its position varies with temperature.

(C) As the electron density is decreased into the insulating phase, the saturation magnetic field $B_{sat}$ decreases and reaches a constant, limiting value $B_{ld}$ at low electron density, with a limiting value $B_{ld}$ that depends on temperature.

As we will argue below, these features are consistent with a model
that considers mixed metallic and insulating phases near the
metal-insulator transition provided that the insulating component is
the phase with the lower electron density. Indeed, local
measurements of the compressibility by Ilani {\it et al.}
\cite{Ilani} in the interacting GaAs-based 2D system have
demonstrated the presence of an admixture of microscopic fractions
of different phases. Suggestions for such mixed phases include an
insulating component associated with a disorder potential
\cite{He1998}, the formation of insulating inclusions due to density
inhomogeneities \cite{DasSarma}, the formation of insulating Wigner
crystallites in a Fermi sea of electrons \cite{SpivakKivelson}, a
non-Fermi-liquid two-phase state involving non-conducting spin
droplets \cite{Teneh,Morgun}, and mesoscopic fluctuations \cite
{Beanninger}.
\subsection{High-density region}

At high densities the two-dimensional electron gas is in a
Fermi-liquid phase and the saturation magnetic field corresponds to
the Zeeman field that fully polarizes the spin of all the electrons
corresponding to a Zeeman field that is of the order of the Fermi
energy. As a result, the saturation magnetic field at high densities
varies linearly with the total density. As the electron density is
reduced toward the kink, regions of an insulating phase begin to
form. Since the insulating regions have density lower than the
remaining Fermi-liquid regions (with correspondingly lower
saturation field), their density will be smaller than the average
density. The density of the remaining Fermi-liquid regions will
therefore be larger than the average density (corresponding to
higher saturation field). As discussed explicitly in Appendix B, the
conductivity of the sample is determined mainly by the properties of
parallel paths through the Fermi-liquid which have much higher
conductance than the parallel paths through insulating regions. As a
result, $B_{sat}$ is determined largely by the conduction through
the Fermi-liquid regions. Since the electron density of Fermi-liquid
regions is higher than the average density, $B_{sat}$ shows an
upward deviation as the average density is decreased.

\subsection{The KINK - A transition by percolation}

The upturn from linear dependence continues  until the percolation
limit is reached. At this point, due to the increase of the volume
fraction of insulating regions, there no longer exists an
uninterrupted path for the electrons to travel through the
Fermi-liquid regions. All the paths connecting the two edges through
the Fermi liquid are now blocked by insulating regions and the (much
lower) conduction proceeds by hopping \cite{Shiqi2017}. At these
densities, the Fermi-liquid and insulating regions act as
resistances in series, and the conductivity of the sample is thus
mainly determined by the lower conductivity of the insulating
regions. The insulating regions have much lower saturation magnetic
fields and the percolation transition thus leads to a sharp drop of
the saturation magnetic field.

In a nutshell, the conduction at densities above the kink depends
overwhelmingly on the conductivity of the Fermi-liquid regions which
have higher than average density, while the conduction below the
kink depends mainly on the conductivity of the insulating regions,
which have lower than average density. Since the conductivity of the
Fermi-liquid and insulating regions affect the density in
fundamentally different ways, the saturation magnetic field measured
through conductance shows the abrupt kink in Fig.~\ref{fig:Fig4}.

The kink occurs at finite temperature, its density, $n_k$, changes with temperature, and $n_k$ is not the same as either the critical density $n_c$ or its value in a magnetic field. In fact, the electron density at which the kink occurs varies with temperature; we expect that its position will depend on the temperature dependence of the volume distribution of insulating and metallic regions. A major implication is that this transition is not the same as the famous quantum phase transition that has been claimed on the basis of numerous measurements in many different dilute two-dimensional strongly interacting systems \cite{review1,review2,review3,Mokashi2012}. Instead, we suggest it is a percolation transition that occurs within a mixed phase system.

It should be noted that a percolation transition has been claimed by
a number of investigators, including He and Xie \cite{He1998} who
proposed a percolation transition due to the disorder landscape
below a liquid-gas critical temperature; Meir \cite{meir1999} who
suggested a percolation transition due to finite dephasing time at
low temperature; and Das Sarma and co-workers \cite
{DasSarma,sarma2007} who advocated a density inhomogeneity driven
percolation transition due to the breakdown of screening in the
random charged impurity disorder background.

\subsection{Low-density region}

Clearly shown in Fig.~\ref{fig:Fig4}, the saturation field $B_{sat}$
decreases as the density is reduced below $n_k$ and becomes constant
at a temperature-dependent low-density value $B_{ld}$ that decreases
with decreasing temperature. The inset to the figure shows the
low-density $B_{ld}$ versus temperature $T$. Based on the few points
available, $B_{sat}$ goes linearly to zero at $T=0$. By the same
token, Fig.~\ref{fig:Fig6} shows that $B_{sat}$ is a linear function
of temperature that is consistent with $B_{sat} \rightarrow 0$ at
$T=0$. As noted earlier, more densely spaced data taken at lower
temperatures are needed to determine whether $B_{sat}$ tends to zero
in the limit of zero temperature.  The slope of these curves
decreases with decreasing density and saturates to a constant value,
as shown in the inset. The fact that $B_{sat}$ versus temperature
becomes independent of the electron density at very low densities
implies that in this regime the magnetoresistance is associated with
non-interacting electrons that respond to the externally applied
magnetic field individually and independently. That the slope of the
curves deviates from this terminal value as the electron density is
increased may be due to the onset of electron-electron interactions.
The limiting value of the slope is the same as the temperature
dependence of saturation magnetic field for the polarization of
single electrons. This feature indicates that the conductivity at
lowest densities is by noninteracting low-density excited electrons
at temperatures well below the Fermi temperature.

\subsection{A quantum phase transition manqu\'ee}

Based on our measurements and analysis of the magnetoconductance of
strongly interacting electrons in 2D in a high-mobility silicon
MOSFET, we suggest that the system is headed toward a  phase
transition between an insulating phase at low density where $B_{sat}
\rightarrow 0$ in the limit of zero temperature and a metallic phase
at higher electron density where $B_{sat}$ remains finite as $T
\rightarrow 0$. However, the system becomes unstable as the critical
point is approached and separates instead into a mixture of
component phases of a nature that has yet to be determined. Rather
than approaching a quantum critical point as the electron density is
varied and the temperature is reduced toward $T=0$, the system
develops mixed conducting/insulating phases where a metal-insulator
transition occurs by percolation.

\subsection{Comparison with earlier work}

The kink we report in this paper was not observed in any of the
earlier magnetotransport measurements of $B_{sat}$ as a function of
electron density
\cite{Yoon2000,Vitkalov2001b,Shashkin2001b,Pudalov2002,tsui2005,Lai2005,popovic,Shlimak2012,Lu2018}.
While a few papers included data for insulating electron densities
\cite{popovic}, most of these earlier measurements were obtained for
densities on the metallic side of the transition. Measurements in
GaAs-based systems by Yoon {\it et al.} \cite{Yoon2000} and in
high-mobility SiGe by Lu {\it et al.} \cite{Lu2018} yielded results
that are in clear disagreement with ours. It is possible that the
results we have obtained require some degree of disorder (not too
much, not too little), and that no mixed phases are formed in the
limit of zero disorder.

Based on measurements of $B_{sat}$ as a function of metallic electron densities in silicon MOSFETs, our group \cite{Vitkalov2001b} and contemporaneous work of Shashkin {\it et al.} \cite{Shashkin2001b} inferred that $B_{sat}$ extrapolates to zero at a finite critical density $n_c$, implying that there is a ferromagnetic instability and a quantum phase transition at that density. By tracing $B_{sat}$ carefully through the transition region, we have now shown that, rather than going to zero as assumed earlier, $B_{sat}$ remains finite (it actually exhibits a kink). This implies that there is no instability and no divergence (at least at finite temperature), and thus no evidence for ferromagnetism in this system, in agreement with recent measurements of Pudalov {\it et al.} \cite{Pudalov2018}.

\section{Summary and Conclusions}

We have shown that the in-plane magnetic field required to fully
polarize the electrons in a strongly interacting electron system in
two dimensions exhibits a clear, heretofore unobserved, kink as a
function of electron density.  On the basis of our data and
analysis, we suggest that the behavior of the 2D electron system in
silicon MOSFETs is consistent with a finite-temperature percolation
transition within a mixed electronic phase.

Our results are novel and important, and promise to open new avenues
of investigation using the behavior of the magnetoconductance as a
tool. In particular, more data at much lower temperatures are needed
to determine the position of the kink, $n_k$, as the temperature,
magnetic field and disorder are varied, and to investigate the
relation of $n_k$ to the critical density, $n_c$, of the
metal-insulator transition determined from measurements of the
resistivity, thermopower, magnetic response, and other experimental
probes.

\section{appendixes}

\subsection{Appendix A}
\vspace{-.4in}
\begin{figure}[h]
\centering
\includegraphics[width=0.6\textwidth]{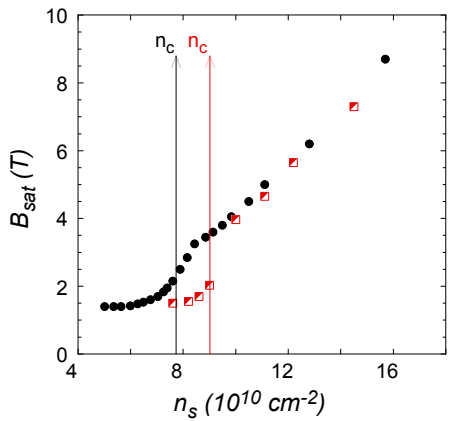}
\vspace{-.7in} \caption{\label{fig:comparison} Comparison between
$B_{sat}$ at $T = 0.27$ K (black dots) for the sample currently
under study and $B_{sat}$ deduced from the conductivity data in
Fig.1(b) in Ref.~\cite{tsui2005}(red open squares); the vertical
lines show the two different critical densities $n_c$ explicitly.}
\end{figure}

It is illuminating to consider a cross comparison between the data
obtained for this sample with similar data obtained earlier for a
sample of lower mobility and a higher level of disorder and higher
critical density $n_c \approx 9.0 \times 10^{10}$ cm$^{-2}$
(compared to $n_c \approx 7.74 \times 10^{10}$ cm$^{-2}$). Shown in
Fig. ~\ref{fig:comparison}, the square (round) symbols denote the
results for the sample with higher (lower) critical density. The
points overlap nicely in the insulating and metallic phase far from
$n_c$. Not surprisingly, however, the behavior of $B_{sat}$ near
$n_c$ is shifted to reflect the different values of the critical
concentration. Clearly, the critical behavior associated with the
metal-to-insulator transition simply shifts as the critical density
shifts to lower densities for lower disorder samples, leaving the
behavior far from the transition intact. This needs to be kept in
mind whenever a comparison is made between different samples.
\subsection{Appendix B}

The general features of the model we propose rely solely on the development with decreasing electron density of a mixed phase composed of a Fermi liquid (or gas) and an insulating component of lower electron density than the liquid. In order to understand the origin of the kink in $B_{sat}$ versus the electron density $n_s$, we need to examine the nature of the conductivity over the entire range of electron densities, both above and below the kink.

Many techniques have been used to calculate the conductivity of composite conductors \cite{perco}. The model we propose below applies to any mixture of a conducting and an insulating phase provided that the insulating phase has the smaller electron density. For specificity, we choose to consider an electronic microemulsion composed of insulating Wigner crystals and a conducting Fermi liquid, as suggested by Spivak and Kivelson \cite{SpivakKivelson, brussarski}. We propose a simplified model that captures the essential effect of the geometry of the crystalline and liquid regions on the measured saturation field.

We consider the sample to be a rectangle with width $D$
perpendicular to the direction of the current and length $L$ along
the direction of current. We assume the effective length of the
Fermi-liquid and Wigner crystal regions are $l_F$ and $l_W$ and
their effective widths are $d_F$ and $d_W$, respectively. At large
densities, $B_{sat}$ decreases linearly with the decrease of
density, the behavior expected for a Fermi liquid.  As the density
is decreased toward $n_k$, the Wigner crystal regions start to
emerge in the parent Fermi liquid. The Wigner crystal regions have
insulating behavior but the sample presents conducting behavior so
long as the Fermi-liquid regions are large enough to percolate
between the two edges. Assuming that the length of the Fermi and
Wigner regions are of the order of the sample length, the
conductance is through parallel Fermi-liquid and Wigner crystal
regions connecting the two ends of the sample. The effective
conductivity of the sample will then be:

\begin{equation}\label{prlc}\sigma= \sigma_F\frac{d_F}{D}+\sigma_W\frac{d_W}{D}\end{equation}
where $\sigma_W$ and $\sigma_F$ are the conductance of Wigner
crystal and Fermi liquid, respectively. Since $\sigma_W\ll\sigma_F$,
and assuming that for densities above $n_k$ the Wigner crystal
regions are not much larger than the Fermi-liquid regions ($d_w
\not\gg d_F$), Eq. (\ref{prlc}) gives $\sigma\approx
\sigma_F\frac{d_F}{D}$ and the normalized conductivity in Eq.
(\ref{eqn:theratio}) reads

\begin{equation}\label{nr}\sigma_{norm}(B)\approx \frac{\Delta \sigma_F}{\Delta\sigma_{F, Total}}=\sigma^F_{norm}(B).\end{equation}

Thus, for densities above $n_k$, the saturation magnetic field corresponds to that of the Fermi liquid. It is important to understand how the emergence of Wigner crystal regions affect the density in the remaining Fermi liquid. Since Wigner crystals have a lower density than the parent Fermi liquid, their formation causes the density of the remaining liquid to increase, with a consequent deviation of $B_{sat}$ upward from straight line behavior as the kink is approached from above. Figures \ref{fig:Fig4} and \ref{fig:Fig5} show such an increase of $B_{sat}$ versus total density $n_s$ as the density is decreased toward the kink.

As the density is decreased further, the fraction of the Wigner
crystal regions increases and, ultimately, there will be no path
through the Fermi-liquid regions to connect the two edges of the
sample. We suggest that the appearance of the kink at the density
$n_k$ corresponds to a percolation transition. At densities below
$n_k$, the paths connecting the two edges of the sample pass through
Wigner crystal and Fermi-liquid regions in series. The effective
conductivity of the sample is then given by
\begin{equation}\sigma=\frac{L}{D}\frac{\sigma_F\sigma_W
d_F}{\sigma_F l_W+\sigma_W l_F}.\label{sr}\end{equation} As the
Wigner crystal and Fermi-liquid regions are in series, their
effective lengths cover the length of the sample, i.e., $L=l_W+l_F$.
The normalized conductivity is then

\begin{equation}\label{sr}\sigma_{norm}(B)=\frac{\frac{\sigma_W(0)}{\frac{\sigma_W(0)}{\sigma_F(0)}+\frac{l_W}{l_F}(0)}-\frac{\sigma_W(B)}{\frac{\sigma_W(B)}{\sigma_F(B)}+\frac{l_W}{l_F}(B)}}{\frac{\sigma_W(0)}{\frac{\sigma_W(0)}{\sigma_F(0)}+\frac{l_W}{l_F}(0)}-\frac{\sigma_W(\infty)}{\frac{\sigma_W(\infty)}{\sigma_F(\infty)}+\frac{l_W}{l_F}(\infty)}}.\end{equation}

Note that, due to the Pomeranchuk effect, the Wigner crystal
fraction over the sample depends on in-plane magnetic field as well
as temperature \cite{SpivakKivelson}. The dependence is explicitly
presented in Eq. (\ref{sr}) as $\frac{l_W}{l_F}(B)$.

To understand the kink in $B_{sat}$ versus the total density $n_s$ curve, we need to examine the conductivity given by Eq. (\ref{sr}) for densities slightly smaller than the kink. In this case, the size of the Wigner crystals is small ($l_W\ll l_F$) such that $\frac{\sigma_W}{\sigma_F}\gg \frac{l_W}{l_F}$ and the difference between the normalized conductivity of the Fermi liquid and the normalized conductivity of the whole sample is approximately:

\begin{equation}\label{dv}\begin{split}
&\frac{\sigma^F_{norm}(B)-\sigma_{norm}(B)}{ \sigma^F_{norm}(B)}\approx\\ &\left[\frac{\sigma_F(0)^2}{\sigma_W(0)}\frac{l_W}{l_F}(0)\left(\frac{1}{\sigma_F(0)-\sigma_F(B)}-\frac{1}{\sigma_F(0)-\sigma_F(\infty)}\right)\right. \\ &-\ \left.\frac{\sigma_F(B)^2}{\sigma_W(B)}\frac{\frac{l_W}{l_F}(B)}{\sigma_F(0)-\sigma_F(B)}+\frac{\sigma_F(\infty)^2}{\sigma_W(\infty)}\frac{\frac{l_W}{l_F}(\infty)}{\sigma_F(0)-\sigma_F(\infty)}\right]
\end{split}\end{equation}

It can be readily seen from Eq. (\ref{dv}) [or Eq. (\ref{sr}) for
that matter] that when $l_W\rightarrow 0$, the normalized
conductivity $\sigma_{norm}(B)$ approaches the normalized
conductivity of the Fermi liquid [$\sigma^F_{norm}(B)$], as was the
case for densities larger than the kink density. This leads to the
continuity of $B_{sat}$ versus density at the kink density $n_k$.

As for the change of the slope of $B_{sat}$ versus density at $n_k$,
we can see from Eq. (\ref{dv}) that, as the density is decreased
below $n_k$, the normalized conductivity starts to deviate from that
of the Fermi liquid.

We now show that $\sigma^F_{norm}(B) > \sigma_{norm}(B)$. As
discussed in Ref. \cite{SpivakKivelson}, at low enough temperature
and for magnetic fields of the order of saturation magnetic field,
the volume fraction of Wigner crystal is of the order of the volume
fraction for infinite magnetic field [i.e., $\frac{l_W}{l_F}(B_c)
\approx \frac{l_W}{l_F}(\infty)$]. Equation (\ref{dv}) then
simplifies to:

\begin{equation}\label{d}\begin{split}
&\frac{\sigma^F_{norm}(B)-\sigma_{norm}(B)}{\sigma^F_{norm}(B)}=\\ & \left[\frac{\sigma_F(0)^2}{\sigma_W(0)}\frac{l_W}{l_F}(0)-\frac{\sigma_F(\infty)^2}{\sigma_W(\infty)}\frac{l_W}{l_F}(\infty)\right]  \times \\ &\left[\frac{1}{\sigma_F(0)-\sigma_F(B)}-\frac{1}{\sigma_F(0)-\sigma_F(\infty)}\right]
\end{split}\end{equation}

Here $\sigma_F(B)$ is a decreasing function of $B$ and its value at
$B=0$ is $2\sigma_F(\infty)$. The conductance of Wigner crystals is
not sensitive to the applied magnetic field. If the
magnetic-field-induced increase of $l_W$ due to the Pomeranchuk
effect is not too large [i.e.,
$\frac{l_W}{l_F}(\infty)<4\frac{l_W}{l_F}(0)$], which we expect at
least for low enough temperatures \cite{SpivakKivelson}, the value
given by Eq.(\ref{dv}) is positive. As a result,
$\sigma_{norm}(B)=0.99$ is achieved at a magnetic field which is
less than the saturation magnetic field of the Fermi liquid. This
result explains the rapid decrease of $B_{sat}$ versus density at
$n_k$ which corresponds to a percolation transition in the mixed
phase.

At the lowest densities, the insulating phase covers the whole
sample. As a result, the conduction is by individually excited
electrons which effectively form a low-density gas. In this limit,
the effect of the Zeeman field is to polarize the spin of individual
electrons and we can readily show that the slope of the saturation
magnetic field versus temperature is the same as for individual
electrons.

\section{Acknowledgments}

We are grateful to Teun Klapwijk for providing the split-gate
geometry samples that enabled these experiments. We thank Steve
Kivelson and Vladimir Dobrosavljevic for valuable discussions and
suggestions, and Sergey Kravchenko for his very careful reading of
the manuscript. This work was supported by the National Science
Foundation Grant No. DMR-1309008 and the Binational Science
Foundation Grant No. 2012210. P.G. acknowledges support by NSF
EFRI-1542863 and PSC-CUNY Award, jointly funded by The Professional
Staff Congress and The City University of New York.

\end{document}